\documentclass[journal]{IEEEtran}

\usepackage{amsmath}
\usepackage{amsfonts}
\usepackage{amsthm}
\usepackage{bm}
\usepackage{graphicx}
\usepackage{booktabs}
\usepackage{multirow}
\usepackage{adjustbox}
\usepackage{float}
\usepackage{placeins}
\usepackage{algorithm}
\usepackage{algorithmic}
\usepackage[hyphens]{url}
\usepackage{cite}
\usepackage{xcolor}
\usepackage{enumitem}
\usepackage{indentfirst}
\usepackage{tikz}
\usetikzlibrary{arrows.meta,positioning,shapes.geometric,calc,fit,backgrounds}
\usepackage{caption}
\usepackage{makecell}
\newif\ifhighlight
\highlightfalse  
\newcommand{\added}[1]{\ifhighlight\textcolor{blue}{#1}\else#1\fi}
\newenvironment{addedblock}{\ifhighlight\color{blue}\fi}{\par\color{black}}
\newcommand{\addedcolor}{\ifhighlight\color{blue}\fi}

\title{Evolutionary Factor Searching for Sparse Portfolio Optimization by Large Language Models}

\makeatletter
\renewcommand{\@makefnmark}{%
  \hbox{\textsuperscript{\normalfont\@thefnmark}}}
\makeatother

\renewcommand{\thefootnote}{%
  \ifcase\value{footnote}
  \or\textdagger
  \or*
  \else\arabic{footnote}
  \fi
}

\author{
Jiandong~Chen\textsuperscript{1}%
\thanks{Jiandong Chen and Haochen Luo contributed equally to this work.},
Haochen~Luo\textsuperscript{1,\textdagger},
Yuan~Zhang\textsuperscript{2},
Chen~Liu\textsuperscript{1}%
\thanks{Corresponding authors: Chen Liu and Qingfu Zhang.},
and Qingfu~Zhang\textsuperscript{1,*}
\\[0.5em]
\small
\textsuperscript{1}City University of Hong Kong,
Hong Kong SAR, China
\\
\textsuperscript{2}Shanghai University of Finance and Economics,
Shanghai, China
}

\usepackage{amsmath,amsfonts,bm, amssymb}
\usepackage{amsthm}

\let\emptyset\varnothing

\def\vomega{\boldsymbol{\omega}}

\def\eqref#1{equation~\ref{#1}}

\def\1{\bm{1}}

\def\rmR{{\mathbf{R}}}

\def\vv{{\bm{v}}}

\def\vx{{\bm{x}}}

\DeclareMathAlphabet{\mathsfit}{\encodingdefault}{\sfdefault}{m}{sl}
\SetMathAlphabet{\mathsfit}{bold}{\encodingdefault}{\sfdefault}{bx}{n}

\def\gF{{\mathcal{F}}}

\def\sR{{\mathbb{R}}}

\usepackage{placeins}  
\let\OLDthebibliography\thebibliography
\renewcommand\thebibliography[1]{%
  \OLDthebibliography{#1}%
  \setlength{\parskip}{0pt}%
  \setlength{\itemsep}{0pt plus 0.2pt}%
}

\begin{document}

\maketitle

\begin{abstract}
Sparse portfolio optimization is a fundamental yet challenging problem in quantitative finance. Traditional approaches often use static objectives and thus adapt poorly to dynamic market regimes. In this work, we propose Evolutionary Factor Search, a framework that leverages large language models and evolutionary algorithms to automatically generate and evolve alpha factors for sparse portfolio construction. The framework recasts asset selection as a ranking task guided by the generated factors and uses an evolutionary feedback loop to iteratively refine the factor pool from portfolio performance. To handle the inter-factor redundancy that accumulates during this search, we further introduce a redundancy-aware weight allocation module that combines random-matrix-theory denoising of the factor correlation matrix with regularized quadratic programming, at negligible overhead and without additional tuning. Extensive experiments on four Fama-French benchmarks and three real-market datasets spanning the United States, Hong Kong, and Mainland China equity markets show that the proposed framework outperforms statistical and optimization-based baselines across diverse markets. Ablation studies further validate the importance of prompt composition, factor diversity, and language-model choice. These results highlight language-model-guided evolution as a robust and interpretable paradigm for portfolio optimization under structural constraints.
\end{abstract}

\begin{IEEEkeywords}
Sparse portfolio optimization, large language models, evolutionary algorithms, alpha factor mining, \added{random matrix theory}.
\end{IEEEkeywords}

\section{Introduction}

Sparse portfolio optimization aims to construct a portfolio by selecting at most $m$ assets from $n$ candidates to optimize key performance metrics, such as cumulative return, risk, or risk-adjusted return. Due to the combinatorial nature of the selection constraint (e.g. $\ell_0$-norm) and the non-convexity of most financial objectives, the problem is known to be NP-hard \cite{lin2024a} and lacks efficient closed-form solutions.

To tackle this important yet challenging problem, classical approaches utilize greedy selection, convex relaxation, mixed-integer programming, and sparsity-regularized optimization models~\cite{brodie2009sparse,lai2018short,dai2018generalized,kremer2020sparse,gunjan2023brief,lin2024a} to compute the exact solutions or their approximations under simplifying assumptions. However, they suffer from two critical limitations: (1) the generated investment suggestions lack interpretability and are difficult for non-experts to understand; (2) the algorithms are often sensitive to hyperparameter choices, leading to unstable performance across market regimes.

\begin{figure*}[htbp]
    \centering
    \includegraphics[width=0.8\textwidth]{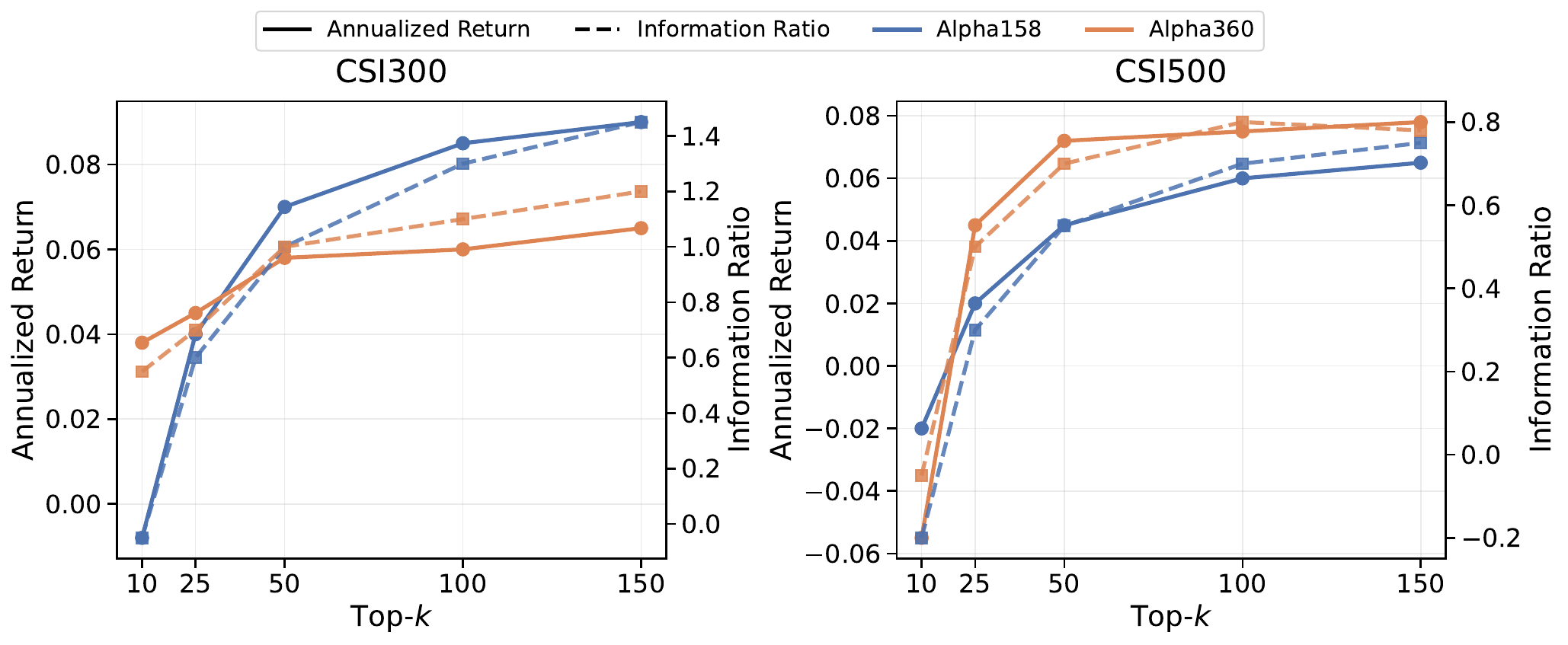}
    \caption{The sparse decay phenomenon: performance of factors from Alpha158 and Alpha360 under different portfolio sparsity ratios in the markets of CSI300 and CSI500 between 2022 and 2025. We report annualized return and information ratio in solid and dashed lines, corresponding to the value on the left and the right axes, respectively.\label{fig:sparse-decay}}
\end{figure*}

To enhance interpretability and adaptability, recent strategies adopt factor-based portfolio construction \cite{ang2014asset,fan2016incorporating}, mapping an asset's historical features (prices, returns, volatility) into a relative-attractiveness score; such mappings are called \textit{alpha factors}. Although more transparent, designing effective factors requires domain expertise, manual tuning, and frequent re-validation; many factors transfer poorly across regimes and decay quickly in live markets. ML-based factor discovery \cite{zhang2020autoalphaefficienthierarchicalevolutionary,yu2023generating} scales poorly to combinatorial asset universes. Moreover, as shown in Fig.~\ref{fig:sparse-decay}, factor libraries from existing methods (e.g., Qlib \cite{yang2020qlibaiorientedquantitativeinvestment}) exhibit \textit{sparse decay} issues: sharp performance drops when selecting 10 or fewer assets, indicating that crafted factors are not precise enough to identify the very best assets under sparse constraints.

Addressing sparse decay requires factor generators that are both \emph{expressive enough} to encode rich market dynamics and \emph{steerable enough} to be guided by performance feedback. Recent advances in large language models (LLMs) make this combination newly attainable: by drawing on strong priors over financial operations and code structures, LLMs can synthesize compact, executable scoring functions that go beyond the symbolic templates of evolutionary or RL-based mining. Indeed, LLMs have already shown impressive capabilities in financial applications, including financial forecasting \cite{zhang2025newquant,li2025llmagentsfinance} and multimodal market analysis \cite{zhang2025newquant,nie2024survey}. These works demonstrate that LLMs can effectively model complex patterns in financial data. However, most LLM-based applications focus narrowly on predictive tasks \cite{yu-etal-2023-harnessing,nie2024survey,zhao2024revolutionizingfinancellmsoverview}, such as price movement classification or sentiment analysis, without addressing the downstream challenges of portfolio construction. Given LLMs' generative nature and ability to synthesize new patterns, recent studies have begun exploring their use in alpha factor discovery \cite{wang2023alphagpthumanaiinteractivealpha,yuan2024alphagpt20humanintheloopai,wang-etal-2024-gpt,li-etal-2024-large-language, Shi_Song_Zhang_Shi_Luo_Ao_Arian_Seco_2025,shi2025navigatingalphajunglellmpowered, tang2025alphaagentllmdrivenalphamining}. While existing research shows the potential of LLMs for generating investment factors, these approaches have two key limitations. First, they rely heavily on human guidance and treat factor mining as a static, one-shot process. This neglects the dynamic nature of financial markets, where alpha signals often decay. Second, studies frequently test these factors on large portfolios of 50 or more assets. This approach overlooks the practical constraints of real-world portfolio management, where factors like implementation costs, risk control, and the need for interpretability necessitate focusing on a much smaller, sparse set of assets.

To address the limitations of traditional alpha mining and the challenges of sparse portfolio optimization, we propose \textbf{E}volutionary \textbf{F}actor \textbf{S}earch (\textbf{EFS}), a novel framework that leverages LLMs and evolutionary algorithms to autonomously generate, evolve, and select alpha factors under $\ell_0$ constraints. EFS integrates LLM-driven creativity with rigorous quantitative feedback, iteratively synthesizing and refining factors based on portfolio-level performance. Unlike prior approaches that rely on static libraries or intermediate ML aggregation, EFS directly produces a single, interpretable scoring function, enabling transparent end-to-end factor mining. By reframing sparse portfolio optimization as an LLM-guided top-$m$ ranking task, EFS inherently aligns with real-world constraints such as limited capital, risk control, and interpretability. Empirical results show EFS achieves state-of-the-art performance against leading benchmarks, and ablation studies highlight the importance of prompt design, search strategy, and redundancy-aware weighting (RW).
\added{However, as the evolutionary search progresses, the factor pool inevitably accumulates structurally similar variants produced by mutation and crossover, leading to inter-factor redundancy. Na\"ively assigning equal weight to all factors fails to exploit this structure. To address this, we further introduce a redundancy-aware factor weight allocation module that automatically identifies and downweights redundant factors, forming a natural complement to the evolutionary factor generation.}

\begin{addedblock}
The primary contributions of this work include:
\begin{enumerate}[label=\arabic*),leftmargin=0pt,itemindent=2em,labelsep=0.4em]
    \item We introduce Evolutionary Factor Search (EFS), an autonomous framework that unifies LLM-driven creativity with rigorous quantitative evaluation. EFS iteratively synthesizes, backtests and refines alpha factors, using performance feedback to guide an evolutionary search for novel, high-performing strategies.
    \item Our framework generates a single, monolithic scoring function, creating a transparent, end-to-end solution for alpha mining. By directly producing a final asset scoring function, EFS eliminates the common need for intermediate machine learning models to aggregate signals, resulting in a more streamlined and interpretable process.
    \item We reframe sparse portfolio optimization as an LLM-guided asset ranking task. Instead of merely identifying factors, EFS produces a direct top-$m$ ranking of assets. The evolutionary search algorithm and the LLM module enable our framework to inherently address real-world constraints like risk control, limited capital, and the need for interpretable, sparse portfolios.
    \item We demonstrate state-of-the-art performance against leading quantitative benchmarks. Furthermore, extensive ablation studies characterize the contribution of key framework components, including our prompt engineering, search depth, and redundancy-aware weight allocation mechanism.
    \item To handle the inter-factor redundancy inherent in evolutionary search, we propose a factor weight allocation module that applies RMT-denoised correlation estimation and regularized quadratic programming to balance predictive power and factor diversity. All hyperparameters of the module are pinned to fixed defaults reused across every dataset and LLM backbone (no per-dataset tuning), so the module integrates seamlessly into the EFS pipeline without per-dataset calibration.
\end{enumerate}
\end{addedblock}

\section{Related Works}

\noindent \textbf{Sparse Portfolio Optimization.} Sparse portfolio optimization aims to balance return, risk, and sparsity. One line of research adds $\ell_0$-regularization to the Markowitz framework to promote sparsity and improve stability \cite{witt1979markowitz,brodie2009sparse,fastrich2015constructing}, while another focuses on index tracking under strict $\ell_0$-constraints to control asset selection and tracking error \cite{li2022sparse}. Other approaches apply ADMM for short-term sparse portfolios in high-frequency settings \cite{lai2018short}, or use structured sparsity like SLOPE \cite{kremer2020sparse}. Recent methods adopt unified indicator relaxations and proximal solvers for mean-CVaR optimization \cite{pmlr-v235-lin24w}, and global solvers for Sharpe maximization under cardinality constraints \cite{lin2024a}. However, these methods remain tied to fixed numerical formulations and often lack adaptability to changing markets.

\noindent \textbf{Alpha Factor Mining and Combination.} Alpha factor mining has progressed significantly with machine learning. Early methods utilized evolutionary algorithms~\cite{zhang2020autoalphaefficienthierarchicalevolutionary} and reinforcement learning~\cite{yu2023generating} to generate and refine factors, forming the basis of a unified framework in~\cite{Shi_Song_Zhang_Shi_Luo_Ao_Arian_Seco_2025}. However, these approaches often lack flexibility and involve complex engineering, limiting their capacity to produce expressive and adaptive factors. Recent advances explore large language models (LLMs) to automate and enhance this process. The Alpha-GPT series~\cite{wang2023alphagpthumanaiinteractivealpha,yuan2024alphagpt20humanintheloopai} introduced human-in-the-loop generation but remained reliant on manual intervention. Later work refined factors using financial signals~\cite{wang-etal-2024-gpt} or symbolic experience chains for better interpretability~\cite{li-etal-2024-large-language}. Most recently, Monte Carlo tree-based search has been employed for more efficient exploration~\cite{shi2025navigatingalphajunglellmpowered}. \added{Concurrent work has further refined LLM-driven alpha mining along three orthogonal axes: AlphaAgent~\cite{tang2025alphaagentllmdrivenalphamining} introduces explicit regularizers to counteract alpha decay; QuantaAlpha~\cite{quantaalpha2026} treats each mining run as a trajectory and applies trajectory-level mutation and crossover; and Cognitive Alpha Mining~\cite{cognitivealpha2025} performs code-based evolution with structured cognitive priors. In parallel, multi-agent and reasoning-based frameworks have emerged for downstream trading: AlphaAgents~\cite{alphaagents2025} and HedgeAgents~\cite{hedgeagents2025} decompose portfolio construction across specialized agents, and Trading-R1~\cite{tradingr1} couples LLM reasoning with reinforcement learning for trade execution.} While promising, these LLM-based approaches still primarily target factor expressivity or trading actions, leaving the question of \emph{how to combine and weight} a redundant pool of LLM-generated factors under sparse cardinality constraints largely untouched; this gap is the focus of EFS.
\added{Beyond factor discovery, the downstream problem of \emph{how to combine} multiple factors and \emph{how to allocate weights} among them remains underexplored. Classical approaches include equal weighting, IC-proportional weighting \cite{grinold2000active}, and mean-variance optimization. A key challenge is covariance estimation: when the number of factors $k$ is non-negligible relative to the sample size $T$, the sample covariance matrix is contaminated by estimation noise \cite{marchenko1967distribution, ledoit2004well}. Random Matrix Theory (RMT) separates signal from noise eigenvalues via the Marchenko-Pastur law and has been widely adopted in financial covariance estimation \cite{bun2017cleaning, laloux1999noise}. We apply RMT denoising to the inter-factor correlation matrix and formulate weight allocation as a regularized quadratic program, complementing the LLM-driven factor generation in EFS.}

\noindent \textbf{Automated Algorithm Design Driven by LLMs.} The factor mining problem shares core traits with automated algorithm design, both involving complex, structured search spaces and a demand for interpretable, high-performing solutions. Recent work has employed LLMs to automate the discovery of algorithms and heuristics for combinatorial optimization. A landmark example is FunSearch~\cite{romeraparedes2024funsearch}, which couples an LLM proposer with an automated evaluator in an evolutionary loop and discovered new mathematical constructions (e.g., for the cap set problem) and improved bin-packing heuristics that surpass the best known human-designed solutions. The Evolution of Heuristics (EoH) framework \cite{liu2023algorithmevolutionusinglarge,pmlr-v235-liu24bs} integrates LLMs with evolutionary computation to generate novel optimization heuristics. Building on this, MEoH \cite{Yao_Liu_Lin_Lu_Wang_Zhang_2025} introduces multi-objective optimization, using a dominance-dissimilarity mechanism to balance performance and efficiency. ReEvo \cite{ye2024reevo} adds reflective evolution, where LLMs iteratively critique and revise their own outputs to improve solution quality. These advances highlight LLMs' capacity to generate adaptive, high-quality heuristics: a capability directly relevant to dynamic and interpretable factor generation in quantitative finance.

\section{Preliminaries}

\subsection{Portfolio Optimization Under the $\ell_0$ Norm Constraint}
\label{sec:prelim_l0}
Portfolio optimization aims to determine the allocation of capital across \( n \) assets to maximize investment performance while controlling risks. The sparse portfolio problem under the \( \ell_0 \)-norm constraint seeks portfolio weights \( \bm{x} \in \mathbb{R}^n \) that optimize a given objective, subject to budget, non-negativity, and sparsity constraints:
\begin{equation}
\begin{aligned} \label{eq:opt-problem}
\max_{\bm{x}} \quad & g(\bm{x})\ \ \text{subject to}\ \  \bm{x}^\top \bm{1} = 1,\ \bm{x} \geq \bm{0},\ \|\bm{x}\|_0 \leq m,
\end{aligned}
\end{equation}
where \( \|\bm{x}\|_0 \) denotes the \( \ell_0 \)-norm of $\bm{x}$, i.e. the number of nonzero entries in \( \bm{x} \), and thus \( m \) specifies the maximum number of assets selected in the portfolio. Without loss of generality, we assume unit total investment, so that the non-negative vector $\bm{x}$ lies in the probability simplex. In a sequential decision-making process of $T$ steps, we may adjust the portfolio weights $\bm{x}$ to approximate the optimality of Problem~(1) at each time stamp $t=0,\dots,T-1$, and we use $\{r_{t+1}\}_{t = 0}^{T-1}$ to represent the corresponding realized total returns. In this context, we use $P_t = \prod_{s = 1}^t r_s$ to represent the cumulative asset value at each time stamp and consider key performance metrics as below:
\begin{itemize}[leftmargin=1.2em]
\item \textit{Cumulative Wealth (CW)} is the total portfolio return ratio, i.e., $\text{CW} = P_T / P_0$, with $P_0 = 1$ by convention so that CW reports the terminal multiple of the initial capital.
\item \textit{Compound Annual Growth Rate (CAGR)} annualizes the cumulative wealth over the test horizon, i.e., $\text{CAGR} = \left(P_T/P_0\right)^{1/Y} - 1 = \text{CW}^{1/Y} - 1$, where $Y$ is the horizon length in years ($Y = T/252$ for daily-return data). CAGR enables a fair comparison across datasets whose test windows differ in length.
\item \textit{Maximum Drawdown (MDD)} is the worst-case loss over a specified period, i.e., $\text{MDD} = \max_{1 \leq i \leq j \leq T} \left( \frac{P_i - P_j}{P_i} \right)$.
\item \textit{Sharpe Ratio (SR)} measures the average return earned in excess of the risk-free rate per unit of volatility. Given the risk-free return $r_f$, it is computed as $\text{SR} = \frac{\bar r_p - r_f}{\text{std}(r_p)}$, where $\bar r_p = \frac{1}{T}\sum_{t=0}^{T-1} r_{t + 1}$ is the average portfolio return, and $\text{std}(r_p)$ is the standard deviation of portfolio returns.
\end{itemize}

The metrics mentioned above are broadly employed in existing literature~\cite{lin2024a, pmlr-v235-lin24w}. An ideal investment policy should have high CW, CAGR, and SR but low MDD.

\subsection{Factor Searching}

In quantitative finance, an alpha factor is a function that assigns a numerical score to each asset based on its historical characteristics and technical indicators such as price, return and volatility. Formally, given an asset with an index \( i \in \{ 1, \dots, n \}\), its historical data over a look-back window of length \( T \) with \( d \) features is represented as a matrix \( \bm{X}_i \in \mathbb{R}^{d \times T} \), where each row corresponds to a different feature (e.g., return, price) and each column corresponds to a time stamp. An alpha factor \( f \) maps this matrix to a scalar score \( f(\bm{X}_i) \), where a higher score implies greater desirability under a specified investment objective (e.g., higher return, lower risk). A typical alpha factor $f$ consists of multiple raw features (e.g. prices, returns), constants and operators. The operators we consider include (1) unary operators (e.g., $\text{abs}(\cdot)$, $\log(\cdot)$); (2) binary operators (e.g., $+$, $-$, $\times$, $/$); (3) time-series operators (e.g., $\text{Sum}(\text{returns}, 5d)$).

An alpha factor can be structured as a computation tree, where leaves are raw features or constants and internal nodes are operators. Our goal is to discover interpretable expressions that generate reliable trading signals that dynamically adapt to the market regime. 
To evaluate the quality of a factor, we use information coefficient (IC) and rank information coefficient (RankIC), which measure the correlation between factor scores and future returns (Pearson and Spearman, respectively). Their time-series stability is quantified by ICIR and RankICIR, calculated as the ratio between the mean and standard deviation of daily IC/RankIC values.

\section{Methodology}

\begin{figure*}[t]
    \centering
    \resizebox{0.95\linewidth}{!}{%
    \begin{tikzpicture}[
        font=\small,
        node distance=3mm and 6mm,
        every node/.style={align=center},
        block/.style={rectangle, rounded corners=2pt, draw=black!70, thick,
                      minimum height=7.5mm, minimum width=20mm, inner sep=2.5pt,
                      fill=blue!6},
        llmblock/.style={block, fill=orange!15, draw=orange!70},
        evalblock/.style={block, fill=green!10, draw=green!55!black},
        weightblock/.style={block, fill=red!8, draw=red!60},
        outblock/.style={block, fill=gray!12, draw=black!60},
        data/.style={cylinder, shape border rotate=90, aspect=0.25,
                     draw=black!70, fill=gray!10, minimum height=7mm, minimum width=14mm},
        arr/.style={-{Latex[length=2mm]}, thick, black!70},
        dashedarr/.style={-{Latex[length=2mm]}, thick, dashed, black!60},
        grouplbl/.style={font=\footnotesize\itshape, text=black!55}
    ]
    \node[data] (market) {Market\\Data};
    \node[data, below=1.5mm of market] (seeds) {Seed\\Factors (TA)};

    \node[llmblock, right=8mm of market] (prompt) {Prompt\\Construction};
    \node[llmblock, right=of prompt, minimum width=24mm] (llm) {LLM\\(Mutation $\oplus$ Crossover)};
    \node[llmblock, right=of llm] (newf) {New Candidate\\Factors $\{f_j\}$};
    \node[evalblock, right=of newf, minimum width=20mm] (eval) {Backtest\\\scriptsize(RankIC, IR)};
    \node[evalblock, right=of eval, minimum width=24mm] (filter) {Filter (Alg.~\ref{alg:factor_filter})\\\scriptsize IC$\geq\!\tau$ $\to$ dedup $\to$ Top-$N$};

    \node[evalblock, anchor=west, minimum width=22mm] (pool) at ($(seeds.east)+(8mm,-7mm)$) {Factor Pool\\$\mathcal{F}_t$};
    \node[weightblock, right=7mm of pool, minimum width=24mm] (corr) {Cross-sectional\\Correlation $\Sigma$};
    \node[weightblock, right=7mm of corr, minimum width=22mm] (rmt) {RMT\\Denoising};
    \node[weightblock, right=7mm of rmt, minimum width=36mm] (qp) {Regularized QP\\$\max\,\boldsymbol{\omega}^\top \mathbf{R} - \lambda\, \boldsymbol{\omega}^\top \widehat{\Sigma}\, \boldsymbol{\omega}$};
    \node[outblock, right=7mm of qp, minimum width=22mm] (port) {Sparse Portfolio\\Top-$m$ Assets};

    \draw[arr] (market.east) -- (prompt.west);
    \draw[arr] (seeds.east) -- ++(3mm,0) |- (prompt.west);
    \draw[arr] (prompt.east) -- (llm.west);
    \draw[arr] (llm.east) -- (newf.west);
    \draw[arr] (newf.east) -- (eval.west);
    \draw[arr] (eval.east) -- (filter.west);

    \draw[arr] (filter.south) |- ($(pool.north)+(0,2mm)$) -- (pool.north);
    \draw[arr] (pool.east) -- (corr.west);
    \draw[arr] (corr.east) -- (rmt.west);
    \draw[arr] (rmt.east) -- (qp.west);
    \draw[arr] (qp.east) -- (port.west);

    \begin{scope}[on background layer]
        \node[fit=(prompt)(llm)(newf), draw=orange!40, dashed, rounded corners,
              inner sep=2mm, label={[grouplbl, orange!70!black]above:LLM-guided Factor Evolution}] {};
        \node[fit=(rmt)(corr)(qp), draw=red!35, dashed, rounded corners,
              inner sep=2mm, label={[grouplbl, red!70!black]below:Weight Allocation (RMT + QP)}] {};
    \end{scope}
    \end{tikzpicture}}
    \caption{
The proposed Evolutionary Factor Search (EFS) framework. Three modules are highlighted: (i)~LLM-guided factor evolution (orange), where top-performing factors from the current pool are used to construct prompts for LLM-based generation of new candidates via mutation and crossover; (ii)~factor pool with backtest evaluation (green); and (iii)~redundancy-aware weight allocation (red), which builds a cross-sectional factor correlation matrix, applies RMT denoising, and solves a regularized quadratic program to assign weights $\boldsymbol{\omega}$ before the top-$m$ assets form the sparse portfolio.
}
    \label{fig:framework}
\end{figure*}

We propose \textbf{Evolutionary Factor Search (EFS)}, an LLM-guided framework that unifies alpha factor discovery with sparse portfolio optimization. Given $n$ assets represented by feature matrices $\{\bm{X}_i\}_{i = 1}^n$, EFS maintains a pool of $k$ alpha factors $\{f_j\}_{j=1}^k$. At each time stamp $t$, each factor $f_j$ produces a raw score vector $\vv_j^{(t)} \in \sR^n$ for $n$ assets; it is then rank-normalized to $\widetilde{\vv}_j^{(t)} \in [0, 1]^n$. We calculate the composite score $s^{(t)} = \sum_{j=1}^k \vomega_j \widetilde{\vv}_j^{(t)}$ by optimized weights $\vomega$, then rank assets, and the top $m$ form the sparse portfolio. By leveraging LLMs' strong priors over financial operations and code structures, EFS directly produces compact, executable scoring functions, eliminating the hand-crafted grammars required by prior evolutionary or RL-based factor miners~\cite{zhang2020autoalphaefficienthierarchicalevolutionary,yu2023generating} and the intermediate ML aggregators used to combine formulaic factors~\cite{Shi_Song_Zhang_Shi_Luo_Ao_Arian_Seco_2025}.

Compared with classical evolutionary methods, EFS employs LLMs as heuristics~\cite{pmlr-v235-liu24bs} when generating new alpha factors.
The LLM serves as a structured generator: prompts encode recent top-performing factors and their performance summaries, and the LLM produces variants via subtree mutation and crossover. 
Compared with handcrafted features and traditional machine learning methods, LLMs are trained on vastly larger corpora and encode much richer heuristics~\cite{nie2024survey,zhang2025newquant}.
Therefore, this single-stage, prompt-controlled pipeline yields end-to-end usability, interpretable evolution, and human-readable factor expressions, making the method well-suited to dynamic financial settings that require rapid iteration and explainability.

In the following subsections, we introduce how EFS generates new alpha factors and how these factors are utilized to generate sparse portfolios.

\subsection{Portfolio Optimization Under Autonomous Factor Search}
\label{sec:efs_method}

The overall framework of EFS is illustrated in Fig.~\ref{fig:framework}. The alpha factor search pipeline proceeds in four stages: (1) factor library warmup, (2) prompt design, (3) iterative evaluation and refinement, and (4) sparse portfolio construction. The first three stages cooperate to evolve a high-quality, non-redundant factor pool, while the fourth stage uses the resulting pool to construct the sparse portfolio. The high-level portfolio construction is described here; the factor weight allocation that further accounts for inter-factor redundancy is elaborated in Section~\ref{sec:weight_allocation}.

\textit{1) Factor Library Warmup.} We begin with a set of seed alpha factors $\{f_1, f_2, \ldots, f_k\}$ as the initial factor library. In practice, we adopt handcrafted popular factor pools such as Alpha101~\cite{kakushadze2016101formulaicalphas} or Alpha158~\cite{yang2020qlibaiorientedquantitativeinvestment}, since the quality of initialization is important in alpha factor search~\cite{anjb-bq76-24}. During warmup, each seed factor is evaluated across an initial look-back window to collect RankIC and cumulative wealth statistics, forming the knowledge base that guides the subsequent LLM-driven search.

\textit{2) Prompt Design.} Prompts are critical when we use LLMs to iteratively improve the quality of alpha factors in the library. In the EFS framework, prompts combine strict task definitions with dynamic feedback. The system prompt defines the role, formatting, constraints, and valid transformation actions (e.g., mutation, crossover). The user prompt provides recent top-performing factors and their anonymized performance metrics (e.g., RankIC, Sharpe) without exposing raw data such as tickers or timestamps. Full prompt templates are in the supplementary material.

\textit{3) Iterative Evaluation and Refinement.} The core module of EFS employs evolutionary algorithms with LLMs in the loop. As the portfolio is rolled forward in time, we define a search frequency $S$ (typically weekly) and conduct the following within each interval. \emph{(a) Factor Generation and Validation:} EFS prompts the LLM to generate new alpha factors, which are formulated as Python functions so that structural and logical correctness can be verified by direct execution; any factor that triggers an execution error is automatically discarded, and the validated factors are evaluated on the look-back window. \emph{(b) Library Update and Pruning:} the factor library is updated by the newly validated factors via Algorithm~\ref{alg:factor_filter}, which first filters candidates using minimum RankIC and RankICIR thresholds (adaptively relaxed if necessary), then enforces diversity by applying hierarchical clustering with Ward's linkage on the inter-factor correlation matrix and retaining only the best-performing factor within each cluster, and finally caps the pool at the top-$N_{\text{top}}\!=\!80$ factors by RankIC.

\begin{algorithm}[htb]\addedcolor
\caption{Factor Filtering Pipeline\label{alg:factor_filter}}
\begin{algorithmic}[1]
\REQUIRE Alpha factor library $\mathcal{C}$ with per-factor metrics  like RankIC, RankICIR and their correlation matrix $\Sigma$; thresholds $\tau_{\text{IC}}\!=\!0.02$, $\tau_{\text{ICIR}}\!=\!0.5$, $\rho_{\max}\!=\!0.7$; minimum pool size $N_{\min}\!=\!8$; top-$N$ cap $N_{\text{top}}\!=\!80$
\ENSURE Filtered pool $\mathcal{F}$
\item[] \textbf{Effectiveness filter}
\STATE $\mathcal{F} \leftarrow \{f_j \in \mathcal{C} : |\text{RankIC}_j| \geq \tau_{\text{IC}}\,\wedge\,|\text{RankICIR}_j| \geq \tau_{\text{ICIR}}\}$
\item[] \textbf{Adaptive relaxation}
\WHILE{$|\mathcal{F}| < N_{\min}$ \AND $\tau_{\text{IC}} > 0.01$}
    \STATE $\tau_{\text{IC}} \leftarrow \max(0.01,\,0.8\,\tau_{\text{IC}})$;\; $\tau_{\text{ICIR}} \leftarrow \max(0.3,\,0.8\,\tau_{\text{ICIR}})$
    \STATE Re-apply Step~1 with relaxed thresholds
\ENDWHILE
\item[] \textbf{Hierarchical-clustering deduplication}
\STATE Calculate the empirical correlation matrix $\Sigma$.
\STATE The inter-factor distance matrix: $\forall i, j, $ $D_{ij} \leftarrow 1 - |\Sigma_{ij}|$.
\STATE Initialize singleton partition $\mathcal{P} \leftarrow \{\{f_1\},\dots,\{f_{|\mathcal{F}|}\}\}$ with $n_C\!=\!1$ for all $C\!\in\!\mathcal{P}$
\WHILE{$|\mathcal{P}| > 1$ \AND $\min_{A\neq B\in\mathcal{P}} D(A,B) \leq 1-\rho_{\max}$}
    \STATE Select: $(A^*,B^*) \leftarrow \arg\min_{A\neq B\in\mathcal{P}} D(A,B)$
    \STATE Merge: $C^* \leftarrow A^* \cup B^*$,\; $n_{C^*} \leftarrow n_{A^*}\!+\!n_{B^*}$
    \item[] \textbf{Lance--Williams update} (Ward's criterion)
    \STATE $\forall E\in\mathcal{P}\setminus\{A^*,B^*\}$, $D(C^{\!*}\!,E) \,\leftarrow\,$
    \begin{equation*}
    \scriptstyle \frac{n_{A^*}\!+n_E}{n_{C^*}\!+n_E}\,D(A^{\!*}\!,E) + \frac{n_{B^*}\!+n_E}{n_{C^*}\!+n_E}\,D(B^{\!*}\!,E) - \frac{n_E}{n_{C^*}\!+n_E}\,D(A^{\!*}\!,B^{\!*})
    \end{equation*}
    \STATE $\mathcal{P} \leftarrow (\mathcal{P}\setminus\{A^*,B^*\})\cup\{C^*\}$
\ENDWHILE
\FORALL{cluster $C \in \mathcal{P}$}
    \STATE Retain $\arg\max_{f_j \in C}\,\text{RankIC}_j$;\; and drop the rest.
\ENDFOR

\item[] \textbf{Top-$N$ cap} (optional)
\IF{$N_{\text{top}}$ is specified \AND $|\mathcal{F}| > N_{\text{top}}$}
    \STATE keep top $N_{\text{top}}$ by RankIC and drop the rest.
\ENDIF
\RETURN $\mathcal{F}$
\end{algorithmic}
\end{algorithm}

\begin{addedblock}
\noindent\textit{Parameter interpretation.} The filtering pipeline is governed by five interpretable controls. The two \emph{effectiveness} thresholds screen each factor's predictive signal: $\tau_{\text{IC}}\!=\!0.02$ is the minimum absolute RankIC, i.e., a factor must rank assets with at least a $2\%$ (Spearman) correlation to next-period returns, below which the signal is essentially indistinguishable from noise at our sample sizes; $\tau_{\text{ICIR}}\!=\!0.5$ is the minimum RankICIR (the ratio of the mean to the standard deviation of the factor's per-period IC), so a factor's average signal must be at least half as large as its temporal fluctuation, ruling out factors whose predictive power is unstable across time. When too few factors clear these bars, the adaptive loop relaxes them geometrically (by $0.8\times$ per round) down to floors $0.01$ and $0.3$ until at least $N_{\min}\!=\!8$ factors survive, guaranteeing a workable pool even in thin markets. The \emph{redundancy} threshold $\rho_{\max}\!=\!0.7$ caps the absolute correlation tolerated between retained factors: any cluster whose members exceed it is collapsed to its single best-RankIC representative, keeping the pool diverse. Finally, the \emph{capacity} cap $N_{\text{top}}\!=\!80$ bounds the pool size to limit downstream computational cost while preserving the strongest factors. These defaults are held fixed across all datasets.
\end{addedblock}

\textit{4) Sparse Portfolio Construction.} Given the refined factor pool, for each trading day we rank assets by the composite score $s^{(t)} = \sum_{j=1}^{k} \vomega_j \cdot \widetilde{\vv}_j^{(t)}$, where $\vomega \in \Delta^{k-1}$ are \emph{factor} weights and $\widetilde{\vv}^{(t)}$ are the rank-normalized scores of each factor; the top-$m$ ranked assets form the sparse portfolio. Capital is then allocated across the selected assets via equal weighting $\vx_i\!=\!1/m$ or positive-score weighting $\vx_i\!=\!\max\{s_i,0\}/\sum_{j}\max\{s_j,0\}$. Daily portfolio returns are tracked across multiple benchmarks for robustness. In the base EFS we use equal factor weights $\vomega_j\!=\!1/k$ as the default; the adaptive scheme that downweights redundant factors is introduced in Section~\ref{sec:weight_allocation}.


\begin{addedblock}
\subsection{Factor Weight Allocation}
\label{sec:weight_allocation}

As described above, EFS iteratively generates new factors
via LLM-driven mutation and crossover. While this process effectively discovers high-quality factors, it also introduces structural redundancy: mutated variants often share similar predictive signals, and the size of the factor library grows over generations.
Equal weighting across all factors clearly fails to account for this redundancy.
Therefore, we adopt an adaptive weight allocation scheme which consists of two steps: (1) we employ random matrix theory (RMT) to denoise and stabilize the inter-factor correlation matrix
$\Sigma$ to obtain $\widehat{\Sigma}$, which ensures positive semidefiniteness of $\widehat{\Sigma}$ and facilitates optimization; (2) we introduce a regularization term $-\lambda\,\boldsymbol{\omega}^\top\widehat{\Sigma}\,\boldsymbol{\omega}$ to perform ``soft deduplication'' by downweighting correlated factors.

Consider the pool of $k$ alpha factors $\gF = \{f_1, \dots, f_k\}$
produced by EFS; at each time $t$, factor
$f_j$ outputs a raw score vector $\vv_j^{(t)} \in \mathbb{R}^n$ across
$n$ assets. Since different factors may operate on different numerical
scales, we first apply cross-sectional rank normalization:
\begin{equation}
    \widetilde{\vv}_j^{(t)} = \text{CrossSectionRank}\!\left(\vv_j^{(t)}\right) \in [0, 1]^n,
    \label{eq:rank_norm}
\end{equation}
and form the composite asset score as a weighted sum:
\begin{equation}
    s^{(t)} = \sum_{j=1}^{k} \vomega_j \cdot \widetilde{\vv}_j^{(t)},
    \quad \vomega \in \Delta^{k-1}
    = \{\vomega \in \mathbb{R}^k_{\geq 0} : \|\vomega\|_1 = 1\}.
    \label{eq:factor_combination}
\end{equation}
The baseline for setting the weights is equal weighting, i.e., $\forall j, \vomega_j = 1 / k$. However, equal weighting does not take the factor quality and diversity into consideration.
The remainder of this section derives an optimized $\vomega^*$ that accounts for factor quality and inter-factor redundancy.


To capture redundancy, we define the cross-sectional Spearman rank
correlation matrix $\boldsymbol{\Sigma} \in \mathbb{R}^{k \times k}$:
\begin{equation}
    \Sigma_{ij} = \frac{1}{T}\sum_{t=1}^{T}
    \rho_s\!\left(\vv_i^{(t)},\,\vv_j^{(t)}\right),
    \label{eq:corr_def}
\end{equation}
where $\rho_s$ denotes Spearman rank correlation. High off-diagonal entries indicate that two factors produce similar asset rankings, causing redundancy.

The empirical correlation $\Sigma$ is contaminated by estimation noise when the
ratio $q = k/T$ is non-negligible. The generalized
Marchenko-Pastur (MP) law~\cite{marchenko1967distribution} states that
under unit-variance noise, the noise contribution concentrates on
eigenvalues falling within the interval
$[(1-\sqrt{k/T})^2,\,(1+\sqrt{k/T})^2]$. We treat eigenvalues exceeding
$\lambda_{\max} = (1+\sqrt{k/T})^2$ as signals, compress the noise
eigenvalues to their mean to preserve the trace, and normalize to unit diagonals to obtain the denoised correlation matrix $\widehat{\Sigma}$. The full procedure is given in Algorithm~\ref{alg:rmt_denoise}.
In our experiments, $q$ ranges from $0.02$ to $0.15$, and $60$--$80\%$ of eigenvalues fall below the MP threshold, indicating significant noise contamination. In practice, the denoising reduces the
condition number of the factor correlation matrix by one to several orders of magnitude (from $2899$ to $253$ on US50, $3.1\!\times\!10^{5}$ to $2897$ on HSI45, and $2.0\!\times\!10^{6}$ to $2799$ on CSI300) while keeping it positive semi-definite, which stabilizes the downstream quadratic program.

\begin{algorithm}[htb]\addedcolor
\caption{RMT Marchenko-Pastur Denoising\label{alg:rmt_denoise}}
\begin{algorithmic}[1]
\REQUIRE Correlation matrix $\Sigma \in \mathbb{R}^{k \times k}$,
  factor count $k$, sample size $T$
\ENSURE Denoised matrix $\widehat{\Sigma}$
\STATE $V, \Lambda \leftarrow \text{eigh}(\Sigma)$
\STATE $\lambda_{\max} \leftarrow (1 + \sqrt{k/T})^2$
\STATE $\mathcal{N} \leftarrow \{i : \lambda_i < \lambda_{\max}\}$,
  $\mathcal{S} \leftarrow \{i : \lambda_i \geq \lambda_{\max}\}$
\IF{$\mathcal{N} = \emptyset$ \OR $\mathcal{S} = \emptyset$}
    \RETURN $\Sigma$
\ENDIF
\STATE $\bar{\lambda} \leftarrow \text{mean}(\{\lambda_i : i \in \mathcal{N}\})$
\FORALL{$i \in \mathcal{N}$}
    \STATE $\lambda_i \leftarrow \bar{\lambda}$
\ENDFOR
\STATE $\widetilde{\Sigma} \leftarrow
  V\,\text{diag}(\lambda_1,\dots,\lambda_k)\,V^\top$
\STATE Normalize: $\widehat{\Sigma}_{ij} \leftarrow
  \widetilde{\Sigma}_{ij} / \sqrt{\widetilde{\Sigma}_{ii}\widetilde{\Sigma}_{jj}}$
\RETURN $\widehat{\Sigma}$
\end{algorithmic}
\end{algorithm}


Besides diversity, we measure each factor's effectiveness via the time-averaged RankIC:
\begin{equation}
    \forall f_j \in \gF,\quad
    R_j = \text{RankIC}(f_j) = \frac{1}{T}\sum_{t=1}^{T}
    \rho_s\!\left(\vv_j^{(t)},\,\mathbf{r}^{(t+1)}\right),
    \label{eq:rankic_def}
\end{equation}
where $\mathbf{r}^{(t+1)} \in \mathbb{R}^n$ is the vector of realized
asset returns on day $t\!+\!1$. The quality vector
$\mathbf{R} = [R_1, \dots, R_k]^\top$ encodes the predictive power of
all factors in the library.

Taking both the denoised matrix $\widehat{\Sigma}$ and the quality vector $\mathbf{R}$ into consideration, we solve the regularized quadratic program:
\begin{equation}
    \max_{\vomega \in \Delta^{k-1}}\;
    \vomega^\top \mathbf{R} - \lambda\,\vomega^\top \widehat{\Sigma}\, \vomega,
    \label{eq:qp_main}
\end{equation}
where $\lambda \geq 0$ balances predictive power and diversity. The first term aims to maximize weighted RankIC, while the second penalizes redundancy by imposing a quadratic cost on correlated factor pairs. Since
$\widehat{\Sigma}$ is positive semidefinite after RMT cleaning, the problem is convex and solved via SLSQP. After obtaining the optimal $\vomega$, we can then calculate the composite asset score by Eq.~(\ref{eq:factor_combination}).

When $\lambda\!=\!0$ with identical RankICs, the optimizer recovers
equal weighting $\vomega_j\!=\!1/k$; as $\lambda$ increases, it
performs ``soft deduplication'' by downweighting redundant factors.
RMT denoising is critical for large factor pools, where the raw
correlation matrix is severely ill-conditioned; after denoising,
the condition number drops by one to several orders of magnitude, enabling
stable QP solutions.
We fix $\lambda\!=\!0.05$ throughout all experiments. The sensitivity
analysis in the next section confirms stable performance when $\lambda \in [0.05, 1.0]$.
In practice, we apply no hyperparameter tuning on $\lambda$ and set $\lambda = 0.05$ for all datasets and LLM choices.
The complete factor weight allocation pipeline is given in Algorithm~\ref{alg:weight_opt}.

\begin{algorithm}[htb]\addedcolor
\caption{Factor Weight Allocation in EFS\label{alg:weight_opt}}
\begin{algorithmic}[1]
\REQUIRE Factor pool $\gF = \{f_1, \dots, f_k\}$, training window
  $[1, T]$, $\lambda = 0.05$
\ENSURE Factor weights $\vomega^*$, composite scores $s^{(t)}$
\item[] \textbf{Factor quality}
\STATE $R_j \leftarrow \text{RankIC}(f_j)$ for each $f_j \in \gF$
\item[] \textbf{Correlation and Denoising}
\STATE $\Sigma_{ij} \leftarrow \frac{1}{T}\sum_{t=1}^{T}
  \rho_s(\vv_i^{(t)}, \vv_j^{(t)})$
\STATE $\widehat{\Sigma} \leftarrow
  \text{RMT\_Denoise}(\Sigma, k, T)$
  \COMMENT{Alg.~\ref{alg:rmt_denoise}}
\item[] \textbf{Solve QP}
\STATE $\vomega^* \leftarrow \arg\max_{\vomega \in \Delta^{k-1}}
  \;\vomega^\top\rmR
  - \lambda\,\vomega^\top\widehat{\Sigma}\vomega$
\item[] \textbf{Weighted scores}
\FOR{each time step $t$}
    \STATE $s^{(t)} \leftarrow \sum_{j=1}^{k}\vomega_j^* \widetilde{\vv}_j^{(t)}$
\ENDFOR
\RETURN $\vomega^*$, $\{s^{(t)}\}$
\end{algorithmic}
\end{algorithm}

\end{addedblock}

\section{Experiments}

\subsection{Experimental Setup}

\begin{table*}[h]
\centering
\footnotesize
\caption{Cumulative Wealth (CW$\uparrow$, terminal multiple $P_T/P_0$), Sharpe Ratio (SR$\uparrow$), and Maximum Drawdown (MDD$\downarrow$) on four Fama-French benchmark datasets (FF25, FF32, FF49, FF100). Arrows indicate the preferred direction of performance. Each FF dataset contains 623 monthly observations (July 1971 to May 2023, $\approx$52 years). Therefore, the absolute CW values reflect the long horizon and can be translated to annualized returns, e.g., the CW of $1 / N$ policy is 374, equivalent to $12.1\%$ annual return; the CW of EFS-GPT-4.1 + RW is 1408, equivalent to $15.0\%$ annual return. Bold marks the best value per dataset at each portfolio size $m$.
\label{tab:bench_performance_metrics}}
\begin{tabular}{@{}ll ccc ccc ccc ccc@{}}
\toprule
& & \multicolumn{3}{c}{FF25} & \multicolumn{3}{c}{FF32} & \multicolumn{3}{c}{FF49} & \multicolumn{3}{c}{FF100} \\
\cmidrule(lr){3-5} \cmidrule(lr){6-8} \cmidrule(lr){9-11} \cmidrule(lr){12-14}
Group & Method & CW$\uparrow$ & SR$\uparrow$ & MDD$\downarrow$ & CW$\uparrow$ & SR$\uparrow$ & MDD$\downarrow$ & CW$\uparrow$ & SR$\uparrow$ & MDD$\downarrow$ & CW$\uparrow$ & SR$\uparrow$ & MDD$\downarrow$ \\
\midrule

\multirow{4}{*}{Baseline}
& 1/N & 374 & 0.23 & 0.55 & 453 & 0.23 & 0.54 & 255 & 0.21 & 0.53 & 389 & 0.21 & 0.55 \\
& SSPO & 77 & 0.14 & 0.78 & 15 & 0.10 & 0.73 & 62 & 0.11 & 0.88 & 1 & 0.05 & 0.82 \\
& Min-CVaR & 387 & 0.24 & 0.54 & 290 & 0.23 & 0.51 & 208 & 0.25 & 0.41 & 156 & 0.20 & 0.54 \\
& Max-Sharpe & 553 & 0.24 & 0.58 & 719 & 0.25 & 0.55 & 211 & 0.23 & 0.42 & 387 & 0.22 & 0.57 \\

\midrule
\multirow{8}{*}{m=10}
& LGBM & 281 & 0.22 & 0.56 & 530 & 0.23 & 0.55 & 169 & 0.19 & 0.59 & 217 & 0.19 & 0.48 \\
& XGB & 292 & 0.22 & 0.52 & 552 & 0.23 & 0.54 & 354 & 0.21 & 0.53 & 392 & 0.20 & 0.50 \\
& mSSRM-PGA & 607 & 0.24 & 0.58 & 749 & 0.25 & 0.55 & 169 & 0.22 & 0.45 & 379 & 0.22 & 0.58 \\
& ASMCVaR & 638 & 0.25 & 0.48 & 670 & 0.24 & 0.45 & 409 & 0.22 & 0.47 & 491 & 0.23 & 0.52 \\
& EFS-DeepSeek-V3 & 640 & 0.25 & 0.51 & \textbf{924} & 0.24 & 0.53 & \textbf{693} & 0.23 & 0.53 & 1233 & 0.24 & 0.55 \\
& ~~+ RW & 1065 & 0.25 & 0.50 & 798 & 0.23 & 0.54 & 441 & 0.19 & 0.59 & 1465 & 0.24 & 0.45 \\
& EFS-GPT-4.1 & 708 & 0.26 & 0.50 & 614 & 0.23 & 0.54 & 565 & 0.22 & 0.54 & 1836 & 0.25 & 0.49 \\
& ~~+ RW & \textbf{1408} & 0.27 & 0.50 & 847 & 0.23 & 0.50 & 593 & 0.20 & 0.61 & \textbf{2435} & 0.26 & 0.51 \\

\midrule
\multirow{8}{*}{m=15}
& LGBM & 313 & 0.22 & 0.55 & 541 & 0.23 & 0.55 & 195 & 0.20 & 0.59 & 224 & 0.19 & 0.49 \\
& XGB & 308 & 0.22 & 0.53 & 491 & 0.22 & 0.55 & 377 & 0.21 & 0.52 & 364 & 0.20 & 0.52 \\
& mSSRM-PGA & 601 & 0.24 & 0.58 & 745 & 0.25 & 0.55 & 172 & 0.22 & 0.45 & 416 & 0.22 & 0.57 \\
& ASMCVaR & 676 & 0.25 & 0.50 & 691 & 0.24 & 0.47 & 527 & 0.24 & 0.45 & 523 & 0.23 & 0.53 \\
& EFS-DeepSeek-V3 & 546 & 0.25 & 0.51 & 716 & 0.24 & 0.54 & 586 & 0.23 & 0.51 & 984 & 0.23 & 0.55 \\
& ~~+ RW & 1050 & 0.25 & 0.50 & 736 & 0.23 & 0.54 & 427 & 0.19 & 0.58 & 1353 & 0.24 & 0.45 \\
& EFS-GPT-4.1 & 530 & 0.25 & 0.52 & 609 & 0.23 & 0.53 & 534 & 0.23 & 0.50 & 1290 & 0.24 & 0.50 \\
& ~~+ RW & \textbf{1365} & 0.27 & 0.50 & \textbf{840} & 0.23 & 0.50 & \textbf{639} & 0.20 & 0.60 & \textbf{2061} & 0.25 & 0.51 \\

\bottomrule
\end{tabular}
\end{table*}

\begin{table*}[h]
\centering
\footnotesize
\caption{Cumulative Wealth (CW$\uparrow$, terminal multiple of initial capital $P_T/P_0$), annualized return (CAGR$\uparrow$), Sharpe Ratio (SR$\uparrow$), and Maximum Drawdown (MDD$\downarrow$) on real-market datasets for $m\!=\!10$. Test windows differ across datasets (US50 covers Jan 2019--Dec 2025, HSI45 Jan 2022--Mar 2025, and CSI300 Jan 2022--May 2025), so CAGR is provided to enable cross-dataset comparison. ``+RW'' denotes the proposed Redundancy-aware Weighting (RW), i.e., factor weights from the RMT-denoised regularized QP of Algorithm~\ref{alg:weight_opt}; rows without ``+RW'' use the default equal factor weighting ($\vomega_j\!=\!1/k$). Bold marks the best value per dataset within each EFS block (gross and net-of-cost); in every case this also coincides with the best value across all methods. The bottom block additionally reports net-of-cost performance under a 10~bp per-trade transaction cost, a commonly cited proxy for typical institutional execution costs in equity trading. Full cost sensitivity ($0/5/10/20$~bp $\times$ $m\!\in\!\{5,10,15,20\}$) is reported in TABLE~\ref{tab:turnover}.\label{tab:market_performance_metrics}}
\begin{tabular}{@{}ll cccc cccc cccc @{}}

\toprule
 & & \multicolumn{4}{c}{US50 (7.0~yr)} & \multicolumn{4}{c}{HSI45 (3.1~yr)} & \multicolumn{4}{c}{CSI300 (3.3~yr)} \\
\cmidrule(lr){3-6} \cmidrule(lr){7-10} \cmidrule(lr){11-14}
Group & Method & CW$\uparrow$ & CAGR$\uparrow$ & SR$\uparrow$ & MDD$\downarrow$ & CW$\uparrow$ & CAGR$\uparrow$ & SR$\uparrow$ & MDD$\downarrow$ & CW$\uparrow$ & CAGR$\uparrow$ & SR$\uparrow$ & MDD$\downarrow$ \\

\midrule
\multirow{3}{*}{Baseline}
& 1/N        &  6.127 & 29.7\% & 1.179 & 0.339 & 1.332 &  9.5\% & 0.436 & 0.409 & 1.083 &  2.5\% & 0.223 & 0.214 \\
& Min-CVaR   &  2.887 & 16.4\% & 1.004 & 0.204 & 1.506 & 13.9\% & 0.633 & 0.362 & 1.389 & 10.6\% & 0.757 & 0.168 \\
& Max-Sharpe & 10.901 & 40.8\% & 1.439 & 0.300 & 1.111 &  3.4\% & 0.262 & 0.377 & 1.088 &  2.6\% & 0.235 & 0.301 \\

\midrule
\multirow{4}{*}{ML / Sparse}
& LGBM       &  4.063 & 22.3\% & 0.853 & 0.373 & 1.313 &  9.0\% & 0.417 & 0.409 & 0.932 & $-$2.1\% & 0.088 & 0.450 \\
& XGBoost    &  4.070 & 22.3\% & 0.862 & 0.351 & 1.227 &  6.7\% & 0.360 & 0.409 & 1.114 &  3.4\% & 0.262 & 0.374 \\
& mSSRM-PGA  & 11.985 & 42.8\% & 1.447 & 0.303 & 1.276 &  8.1\% & 0.397 & 0.388 & 1.000 &  0.0\% & 0.175 & 0.447 \\
& ASMCVaR    &  3.575 & 20.0\% & 1.170 & 0.205 & 1.511 & 14.0\% & 0.636 & 0.361 & 1.464 & 12.4\% & 0.826 & 0.165 \\

\midrule
\multirow{4}{*}{EFS gross}
& DeepSeek-V3           & 10.651 & 40.2\% & 1.437 & 0.321 & 1.794 & 20.4\% & 0.983 & 0.283 & 1.385 & 10.5\% & 0.827 & 0.196 \\
& DeepSeek-V3+RW & \textbf{15.997} & \textbf{48.6\%} & \textbf{1.638} & 0.313 & 1.794 & 20.4\% & 0.983 & 0.283 & 1.376 & 10.3\% & 0.795 & 0.187 \\
& GPT-4.1            & 12.178 & 42.9\% & 1.406 & 0.343 & \textbf{1.941} & \textbf{23.5\%} & \textbf{1.130} & \textbf{0.231} & 1.468 & 12.5\% & 0.958 & 0.159 \\
& GPT-4.1+RW  & 12.116 & 42.8\% & 1.410 & 0.345 & 1.865 & 21.9\% & 1.055 & 0.266 & \textbf{1.511} & \textbf{13.5\%} & \textbf{1.014} & \textbf{0.152} \\

\midrule
\multirow{4}{*}{EFS @ 10bp}
& DeepSeek-V3           &  2.663 & 15.0\% & 0.671 & \textbf{0.387} & 1.700 & 18.4\% & 0.902 & 0.293 & 1.132 &  3.9\% & 0.357 & 0.222 \\
& DeepSeek-V3+RW        &  2.188 & 11.8\% & 0.558 & 0.413 & 1.700 & 18.4\% & 0.902 & 0.293 & 1.064 &  1.9\% & 0.209 & 0.214 \\
& GPT-4.1            & \textbf{3.969} & \textbf{21.8\%} & \textbf{0.839} & 0.431 & \textbf{1.874} & \textbf{22.1\%} & \textbf{1.074} & \textbf{0.234} & \textbf{1.187} & \textbf{5.4\%} & \textbf{0.465} & \textbf{0.190} \\
& GPT-4.1+RW         &  3.103 & 17.6\% & 0.717 & 0.390 & 1.760 & 19.7\% & 0.967 & 0.275 & 1.140 &  4.1\% & 0.367 & 0.197 \\

\bottomrule
\end{tabular}

\end{table*}

\noindent \textbf{Datasets.} We evaluate all methods on both academic benchmarks and real-world asset pools. For academic testing, we use four standard monthly-return datasets from the Kenneth R. French Data Library: \textit{FF25}, \textit{FF32}, \textit{FF49} and \textit{FF100}, with portfolio sizes indicated by their names. Each spans 623 monthly observations from July 1971 to May 2023 ($\approx$52 years). To assess real-world applicability, we construct three daily-return datasets across major markets: \textit{US50} (50 U.S. stocks from 2019--2025), \textit{HSI45} (45 Hang Seng Tech firms from 2022--2025), and \textit{CSI300} (full CSI 300 index from 2022--2025). These datasets span diverse market regimes and test both robustness and scalability. This dual setup ensures our model is validated across theoretical and practical investment scenarios. \added{Every dataset is provided purely as a return series (the Fama-French portfolios and the market pools alike); following common practice we recover a price trajectory as the cumulative product of returns, so the EFS factors are functions of these returns and the derived prices rather than of full OHLCV inputs.} \added{Notably, all three real-market datasets extend through 2025 (US50 through December 2025, HSI45 through March 2025, and CSI300 through May 2025), providing a non-trivial portion of each test horizon that strictly postdates the training cutoffs of the LLM backbones we use (mid- to late-2024); this segment is exploited in the out-of-sample robustness analysis in TABLE~\ref{tab:oos}.}

\noindent \textbf{Methods.} To evaluate the performance of our EFS framework, we benchmark it against a comprehensive suite of established portfolio construction methodologies. These baselines are divided into two main categories: traditional non-sparse portfolio strategies and modern sparse portfolio models. Non-sparse benchmarks include Equal weighting ($1/N$) portfolio, Minimum conditional Value at Risk (Min-CVaR) optimization, Maximum Sharpe Ratio (Max-Sharpe) optimization. Sparse benchmarks include a short-term sparse portfolio optimizer SSPO \cite{lai2018short} (used on FF benchmarks only), general machine learning selectors XGBoost \cite{Chen:2016:XST:2939672.2939785}, LightGBM (LGBM) \cite{ke2017lightgbm}, more recent sparse-portfolio strategies mSSRM-PGA \cite{lin2024a} and ASMCVaR \cite{pmlr-v235-lin24w}.

\noindent \textbf{Evaluation.} We assess portfolio performance using standard metrics: \textit{Cumulative Wealth (CW)}, \textit{Compound Annual Growth Rate (CAGR)}, \textit{Sharpe Ratio (SR)}, and \textit{Maximum Drawdown (MDD)}, as defined in Section~\ref{sec:prelim_l0}. CAGR is reported only for the real-market datasets, whose test windows differ in length. Daily Sharpe Ratios are computed with zero risk-free rate to avoid bias across shifting rate regimes. 

\noindent \textbf{Implementation.} For each market dataset we repeat the factor search three times and pool the discovered factors into a unified library for final backtesting (\textit{aggregated evaluation}), reducing LLM-randomness and market-noise variance. We use two online LLM backbones: GPT-4.1 and DeepSeek-V3, accessed via the \texttt{deepseek-chat} production endpoint at the time of our experiments (the endpoint has subsequently been remapped to the DeepSeek-V4 family released in April 2026). 
As discussed above, we use these earlier version LLMs to avoid data leakage, since they are published before the time range of our real-market test data.
For simplicity and compatibility with LLMs, factor construction uses only closing prices and returns.
\added{
We use equal factor weighting ($\vomega_j = 1/k$) by default; the results based on redundancy-aware weighting in Algorithm~\ref{alg:weight_opt} are reported using the ``+RW'' label, such as the ones in TABLE~\ref{tab:bench_performance_metrics}. When selecting factors to feed the weight allocation, we apply a per-dataset IC screen (IC$\geq$0.05 for HSI45 and CSI300, IC$\geq$0.005 for US50) matched to each market's typical factor-IC magnitude; this factor-selection threshold is a data-preprocessing choice and is separate from the weight allocation module's hyperparameters, which use fixed defaults across all datasets.
}

\subsection{Results of Portfolio Performance}
\label{sec:portfolio_results}

We evaluate our method on both Fama-French academic benchmarks and real-market datasets. Results are reported in TABLE~\ref{tab:bench_performance_metrics} and TABLE~\ref{tab:market_performance_metrics} using standard performance metrics: Cumulative Wealth (CW), Sharpe Ratio (SR), and Maximum Drawdown (MDD).

\noindent \textbf{Benchmark Dataset Results.}
TABLE~\ref{tab:bench_performance_metrics} presents results across four Fama-French benchmark datasets. Under the standard sparse setting ($m\!=\!10$), our proposed EFS framework achieves the best performance across all asset pools, with only minor differences between the GPT-4.1 and DeepSeek-V3 backends, demonstrating the robustness of our LLM-guided evolution process. Notably, the performance gap widens as the dataset size increases, confirming the scalability and advantage of EFS in larger universes (e.g., FF100). Furthermore, incorporating the redundancy-aware weighting (denoted ``+RW'' in TABLE~\ref{tab:bench_performance_metrics}) yields gains in most settings, especially for the GPT-4.1 backbone which improves over its equal-weight counterpart across all four datasets at both $m\!\in\!\{10,15\}$. These results indicate that the EFS framework and redundancy-aware weighting improve capital allocation while preserving sparsity, especially under challenging high-dimensional settings.

\noindent \textbf{Real-Market Dataset Performance.} TABLE~\ref{tab:market_performance_metrics} compares EFS against traditional baselines (1/N, Min-CVaR, Max-Sharpe), ML-based selectors (LGBM, XGBoost), and recent sparse optimization methods (mSSRM-PGA, ASMCVaR). The EFS results include both the default equal-weight scheme and the proposed redundancy-aware (+RW) variant for each backend. Under the equal weighting scheme, EFS-GPT-4.1 already achieves better cumulative wealth than traditional and ML-based baselines across all three datasets while maintaining very competitive Sharpe ratio and maximum drawdown.
The redundancy-aware weighting (+RW) further amplifies the overall performance, especially on US50, DeepSeek-V3+RW reaches CW=16.00 (SR=1.64), substantially above the strongest baseline mSSRM-PGA (CW=11.99, SR=1.45). The redundancy-aware weighting can achieve better or comparable performance in other cases as well.
In addition, the results in TABLE~\ref{tab:market_performance_metrics} demonstrate that our EFS framework is robust to different backend LLMs, as the performance based on both DeepSeek-V3 and GPT-4.1 is competitive. 
Finally, we notice that ML baselines (LGBM, XGBoost) underperform classical portfolio methods, suggesting that supervised prediction of next-day returns from simple features is not competitive with factor-based ranking. 
In the lower block of TABLE~\ref{tab:market_performance_metrics}, we consider the more realistic cases with 10~bp transaction cost (one basis point $=0.01\%$ of traded notional per trade), where the higher turnover induced by redundancy-aware weighting offsets its gross-return advantage, so equal weighting becomes the better choice. Detailed turnover and transaction-cost sensitivity analysis is reported in TABLE~\ref{tab:turnover}, the discussions are deferred to Section~\ref{subsec:analysis}.

\begin{figure}[h]
\centering
\includegraphics[width=\linewidth]{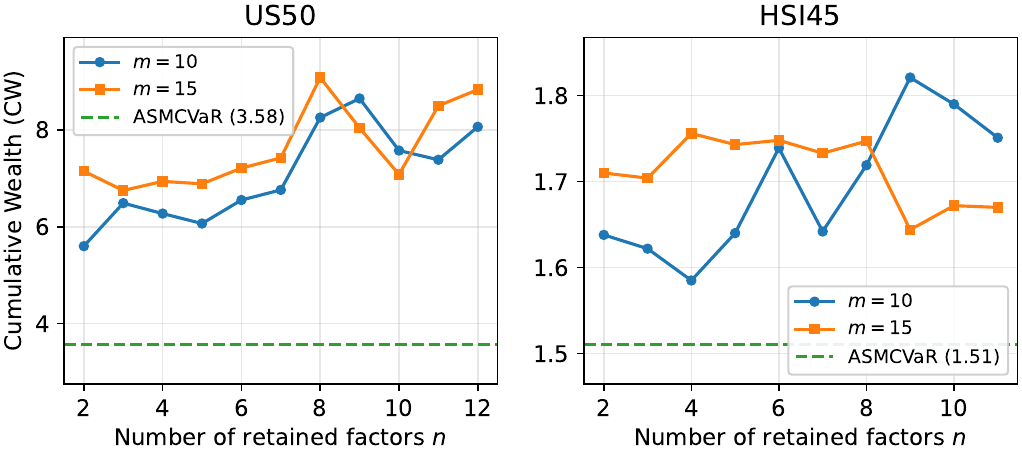}
\caption{Sensitivity of EFS cumulative wealth to the number of top-RankIC factors retained ($n$), equal-weighted, on US50 and HSI45 (DeepSeek-V3 pool). CW stays well above the strongest baseline (ASMCVaR, dashed) across the whole range, and the $m\!=\!15$ portfolio varies less than $m\!=\!10$; the mean $|\text{RankIC}|$ of the retained set is also essentially flat ($\approx\!0.012$ on US50, $\approx\!0.051$ on HSI45).\label{fig:factor_count}}
\end{figure}

EFS is also robust to the number of top-RankIC factors retained for portfolio construction (Fig.~\ref{fig:factor_count}): as the retained count $n$ is varied from $2$ to $10$ on US50 and HSI45, cumulative wealth stays consistently high and well above the strongest baseline ASMCVaR (CW $3.58$ on US50 and $1.51$ on HSI45), the more diversified $m\!=\!15$ portfolio is even less sensitive than $m\!=\!10$, and the mean RankIC of the selected factors remains essentially flat throughout.

\subsection{Ablation Studies}
\label{sec:ablation}

\begin{table*}[h]
\centering
\footnotesize
\caption{Ablation of the weight-allocation module on the three real-market datasets at $m\!=\!10$ (main factor pools, GPT-4.1 backbone), using the same CW/CAGR/SR/MDD criteria as TABLE~\ref{tab:market_performance_metrics}. \textbf{Block~1} ablates the module components, \textbf{Block~2} its regularization strength $\lambda$, and \textbf{Block~3} the factor-pool size $k$. The ablation of the LLM-search inputs is reported separately in TABLE~\ref{tab:ablation_inputs}.} \label{tab:ablation}
\begin{tabular}{@{}cl cccc cccc cccc @{}}

\toprule
 & & \multicolumn{4}{c}{US50 (7.0~yr)} & \multicolumn{4}{c}{HSI45 (3.1~yr)} & \multicolumn{4}{c}{CSI300 (3.3~yr)} \\
\cmidrule(lr){3-6} \cmidrule(lr){7-10} \cmidrule(lr){11-14}
Block & Configuration & CW$\uparrow$ & CAGR$\uparrow$ & SR$\uparrow$ & MDD$\downarrow$ & CW$\uparrow$ & CAGR$\uparrow$ & SR$\uparrow$ & MDD$\downarrow$ & CW$\uparrow$ & CAGR$\uparrow$ & SR$\uparrow$ & MDD$\downarrow$ \\

\midrule
\multirow{3}{*}{\makecell{Block 1: \\ Weight \\ Allocation}}
& Equal Weight & 12.18 & 42.9\% & 1.41 & 0.343 & 1.94 & 23.4\% & 1.13 & 0.231 & 1.47 & 12.5\% & 0.96 & 0.159 \\
& RW & 12.12 & 42.8\% & 1.41 & 0.345 & 1.87 & 22.0\% & 1.06 & 0.266 & 1.51 & 13.5\% & 1.01 & 0.152 \\
& ~w/o RMT & 11.05 & 40.9\% & 1.37 & 0.350 & 1.91 & 22.8\% & 1.11 & 0.270 & 1.49 & 13.0\% & 0.99 & 0.149 \\
\midrule
\multirow{6}{*}{\makecell{Block 2: \\ Regularizer}}
& $\lambda = 0.001$ & 10.49 & 39.9\% & 1.36 & 0.349 & 1.82 & 21.0\% & 1.01 & 0.288 & 1.42 & 11.3\% & 0.88 & 0.184 \\
& $\lambda = 0.01$ & 11.38 & 41.5\% & 1.39 & 0.358 & 1.82 & 21.0\% & 1.02 & 0.287 & 1.49 & 13.0\% & 0.99 & 0.154 \\
& $\lambda = 0.05$ & 12.12 & 42.8\% & 1.41 & 0.345 & 1.87 & 22.0\% & 1.06 & 0.266 & 1.51 & 13.5\% & 1.01 & 0.152 \\
& $\lambda = 0.1$ & 11.23 & 41.3\% & 1.38 & 0.359 & 1.82 & 21.0\% & 1.02 & 0.278 & 1.53 & 13.9\% & 1.04 & 0.149 \\
& $\lambda = 0.5$ & 10.79 & 40.5\% & 1.35 & 0.361 & 1.99 & 24.4\% & 1.18 & 0.254 & 1.53 & 13.9\% & 1.04 & 0.155 \\
& $\lambda = 1$ & 11.06 & 41.0\% & 1.37 & 0.345 & 1.99 & 24.4\% & 1.17 & 0.256 & 1.53 & 13.9\% & 1.04 & 0.154 \\
\midrule
\multirow{5}{*}{\makecell{Block 3: \\ Pool Size}}
& $k = 5$ & 11.70 & 42.1\% & 1.40 & 0.338 & 1.80 & 20.5\% & 1.00 & 0.255 & 1.46 & 12.3\% & 0.94 & 0.197 \\
& $k = 10$ & 11.19 & 41.2\% & 1.36 & 0.352 & 1.88 & 22.2\% & 1.07 & 0.264 & 1.45 & 12.1\% & 0.90 & 0.196 \\
& $k = 20$ & 11.75 & 42.2\% & 1.39 & 0.345 & 1.85 & 21.6\% & 1.05 & 0.245 & 1.53 & 13.9\% & 1.04 & 0.147 \\
& $k = 40$ & 12.04 & 42.7\% & 1.40 & 0.342 & 1.99 & 24.4\% & 1.17 & 0.231 & 1.49 & 13.0\% & 0.99 & 0.155 \\
& $k = 80$ & 12.19 & 42.9\% & 1.41 & 0.343 & \multicolumn{4}{c}{N/A (HSI45 pool $<\!80$)} & 1.47 & 12.5\% & 0.96 & 0.159 \\
\bottomrule
\end{tabular}
\end{table*}

\begin{table*}[t]
\centering
\footnotesize
\caption{Ablation of the LLM-search input components at $m\!=\!10$, using the same CW/CAGR/SR/MDD criteria as TABLE~\ref{tab:market_performance_metrics}. Each variant is an \emph{independent reduced-budget run} ($50$ search iterations on each dataset's held-out window with equal asset weighting and the DeepSeek-V3 backbone), so the values are comparable across variants but on a smaller absolute scale than the main-pool results in TABLE~\ref{tab:ablation}. \emph{w/o Quality} drops factor-level RankIC from the prompt, \emph{w/o Performance} drops portfolio CW feedback, \emph{w/o TA} drops the Alpha101 seeds, and \emph{Initial Factor} skips the LLM-driven search.\label{tab:ablation_inputs}}
\begin{tabular}{@{}l cccc cccc cccc@{}}
\toprule
 & \multicolumn{4}{c}{US50} & \multicolumn{4}{c}{HSI45} & \multicolumn{4}{c}{CSI300} \\
\cmidrule(lr){2-5}\cmidrule(lr){6-9}\cmidrule(lr){10-13}
Configuration & CW$\uparrow$ & CAGR$\uparrow$ & SR$\uparrow$ & MDD$\downarrow$ & CW$\uparrow$ & CAGR$\uparrow$ & SR$\uparrow$ & MDD$\downarrow$ & CW$\uparrow$ & CAGR$\uparrow$ & SR$\uparrow$ & MDD$\downarrow$ \\
\midrule
Full & 1.593 & 12.4\% & 0.87 & 0.369 & 1.354 & 19.5\% & 0.88 & 0.299 & 1.051 & 2.8\% & 0.61 & 0.205 \\
w/o Quality & 1.862 & 16.9\% & 1.08 & 0.264 & 1.154 & 8.8\% & 0.69 & 0.215 & 1.148 & 7.8\% & 0.87 & 0.165 \\
w/o Performance & 1.190 & 4.5\% & 0.61 & 0.313 & 1.231 & 13.0\% & 0.71 & 0.326 & 0.972 & $-$1.6\% & 0.42 & 0.296 \\
w/o TA & 1.637 & 13.2\% & 0.95 & 0.331 & 1.151 & 8.7\% & 0.58 & 0.310 & 0.843 & $-$8.9\% & $-$0.01 & 0.386 \\
Initial Factor & 1.328 & 7.4\% & 0.72 & 0.272 & 1.518 & 27.9\% & 1.05 & 0.320 & 0.956 & $-$2.4\% & 0.39 & 0.314 \\
\bottomrule
\end{tabular}
\end{table*}

\noindent \textbf{Inputs.} To analyze the contribution of the key input components in our EFS framework, we ablate the LLM-search inputs and report the results in TABLE~\ref{tab:ablation_inputs}. Following the same datasets and metrics as TABLE~\ref{tab:market_performance_metrics}, each variant is run on US50, HSI45, and CSI300 at portfolio size $m\!=\!10$ with equal asset weighting and the same DeepSeek-V3 backbone, under an independent reduced-budget search (50 iterations) so that differences in test-phase performance can be attributed to the ablated component.
In this context, we evaluate four ablations: (1) removing factor-quality features such as RankIC from the LLM prompt (w/o Quality); (2) removing portfolio-performance feedback such as CW from the prompt (w/o Performance); (3) removing all technical-analysis (TA) based seed factors (w/o TA); and (4) skipping the LLM-driven expansion altogether and evaluating the initial handcrafted library only (Initial Factor). The Full baseline keeps all components enabled.

Across the three markets the component contributions are consistent in direction for two of the inputs and market-dependent for the others. \emph{Portfolio-performance feedback} is the most consistently useful signal: \emph{w/o Performance} lowers cumulative wealth on every dataset and turns CSI300 from a small gain into a loss ($1.051\!\rightarrow\!0.972$), confirming that portfolio-level feedback steers the search toward profitable, not merely high-IC, factors. \emph{TA seeds} matter most on the Asian markets: \emph{w/o TA} is the weakest variant on HSI45 and collapses CSI300 to a capital loss ($1.051\!\rightarrow\!0.843$, CAGR $-8.9\%$), while barely affecting US50, indicating the seed library supplies a useful inductive prior where the LLM has less price history to exploit. In contrast, \emph{factor-quality (RankIC) feedback} is the one component whose value is mixed: removing it \emph{improves} US50 and CSI300 yet degrades HSI45, suggesting that IC-level guidance can over-emphasize ranking accuracy at the expense of portfolio return once the factor pool is already large, while still helping on the factor-scarce HSI45 pool. Finally, \emph{Initial Factor} (no LLM search) is competitive only on the short HSI45 window, where it even edges out Full, but is clearly weaker on US50 and loses capital on CSI300, showing that the LLM-driven expansion adds the most value on the larger, longer-horizon universes. Because these ablations use a reduced 50-iteration budget and a single run per setting, the per-dataset gaps carry non-trivial run-to-run variance and should be read qualitatively rather than as precise rankings.

\begin{addedblock}
\noindent\textbf{Weight Allocation.}
We ablate the components of our weight allocation module across all three datasets ($m\!=\!10$), as reported in Block 1 of Table~\ref{tab:ablation}. Specifically, we consider three factor weight allocation schemes with the default hyperparameters as in Table~\ref{tab:market_performance_metrics} and using the backbone GPT-4.1: (1) using equal weights (Equal Weight); (2) using redundancy-aware weighting as in Algorithm~\ref{alg:weight_opt} (RW); (3) removing the RMT denoising part in RW (w/o RMT).
Our results indicate that equal weight and RW \textit{without hyperparameter optimization} can achieve comparable performance in terms of all metrics. Notably, when removing RMT, we see considerable performance degradation on US50, indicating that the empirical correlations among different alpha factors discovered on US50 is more noisy.

In Table~\ref{tab:e2e_weight}, we further compare our method against three heuristic factor-weighting schemes under different markets and asset sparsity $m$. Let $\bm{r}$ denote the vector of per-factor RankICs, $\Sigma$ the empirical inter-factor correlation matrix, and $\vomega\in\Delta^{k-1}$ the non-negative factor weights summing to one. The schemes are: (1) \emph{IC-Prop}, with weights proportional to the clipped information coefficient, $\vomega_j \propto \max(r_j,0)$; (2) \emph{Min-Var}, the minimum-variance allocation $\arg\min_{\vomega}\,\vomega^\top\Sigma\vomega$, which ignores the IC signal and only suppresses correlated factors; and (3) \emph{Max-Sharpe}, which maximizes the IC-to-risk ratio $\vomega^\top\bm{r}/\sqrt{\vomega^\top\Sigma\vomega}$. All three share the same simplex constraints as our RW; RW differs by maximizing the RMT-regularized objective $\vomega^\top\bm{r}-\lambda\,\vomega^\top\widehat{\Sigma}\vomega$ on the denoised matrix $\widehat{\Sigma}$ (Algorithm~\ref{alg:weight_opt}), which is what makes it robust to correlation-estimation noise. The results indicate that our proposed RW can achieve the best and stable performance, especially in the sparse portfolio regime, i.e., a small $m$.

Furthermore, the computational overhead introduced by RW is negligible. During each update interval for the alpha factor pools, the RW module requires only 93.5 ms to execute, which includes both the RMT and the quadratic programming (QP) step formulated in (\ref{eq:qp_main}). In contrast, the LLM inferences within the EFS framework take several minutes to complete, representing the primary computational bottleneck of the system.

\begin{table*}[!t]
\centering
\caption{End-to-end cumulative wealth of different factor-weighting schemes across $m\!\in\!\{5,10,15,20\}$.\label{tab:e2e_weight}}
\small
\begin{tabular}{ll cccc cccc cccc}
\toprule
& & \multicolumn{4}{c}{US50} & \multicolumn{4}{c}{HSI45} & \multicolumn{4}{c}{CSI300} \\
\cmidrule(lr){3-6}\cmidrule(lr){7-10}\cmidrule(lr){11-14}
& Method & $m\!=\!5$ & $10$ & $15$ & $20$ & $m\!=\!5$ & $10$ & $15$ & $20$ & $m\!=\!5$ & $10$ & $15$ & $20$ \\
\midrule
& Equal         & 13.27 & 10.65 & 10.48 & 10.11 & 2.03 & 1.79 & 1.65 & 1.69 & \textbf{1.54} & 1.39 & 1.40 & 1.41 \\
& IC-Prop       & 10.05 & 10.10 & 11.75 & 9.80  & 2.01 & 1.76 & 1.63 & 1.68 & \textbf{1.54} & 1.39 & 1.39 & 1.41 \\
& Min-Var       & 19.14 & 15.73 & 13.91 & 12.74 & 1.73 & 1.69 & 1.71 & 1.63 & 1.46 & 1.37 & 1.43 & 1.43 \\
& Max-Sharpe    & 24.87 & 17.31 & 15.35 & 15.23 & 1.74 & 1.69 & 1.78 & 1.58 & 1.53 & 1.40 & 1.46 & 1.50 \\
& RW & \textbf{25.32} & 16.00 & 12.94 & 13.88 & \textbf{2.03} & 1.79 & 1.65 & 1.69 & 1.49 & 1.38 & 1.41 & 1.50 \\
\bottomrule
\end{tabular}
\end{table*}

\begin{figure*}[htbp]
    \centering
    \includegraphics[width=0.7\textwidth]{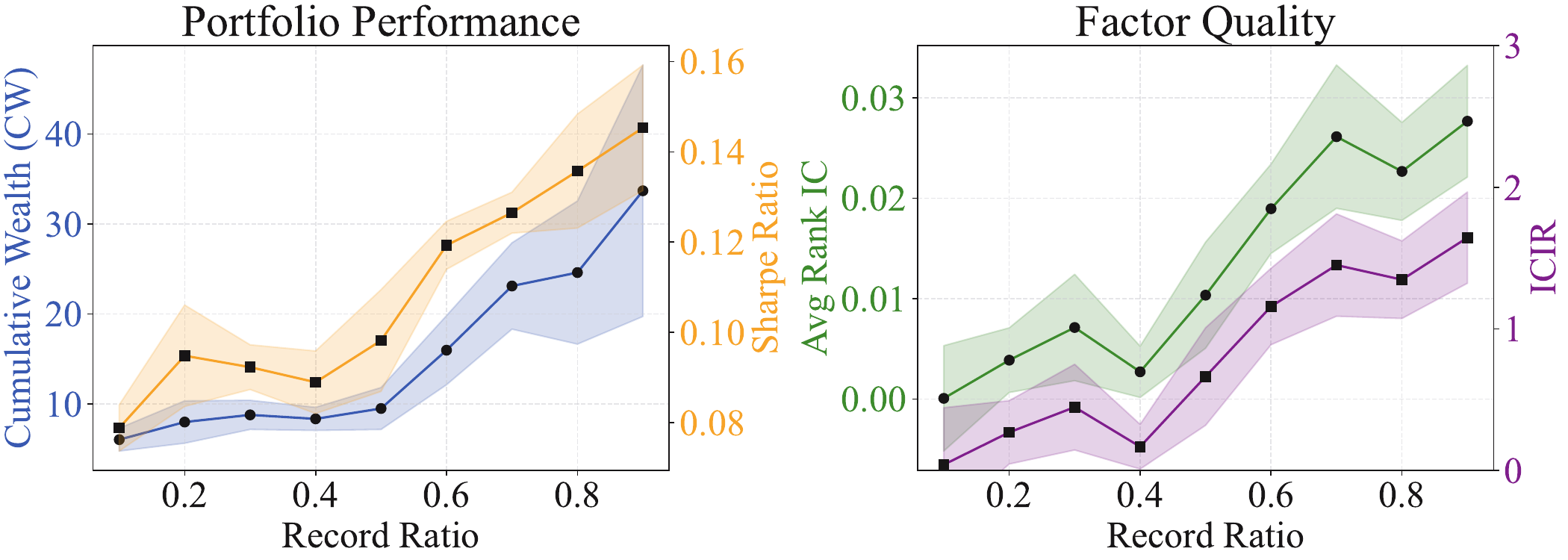}
    \caption{Effect of warm-up duration on EFS performance (US50). Each point freezes the factor pool at a different fraction of the full evolutionary search progress (10\%--90\%) and reports the resulting cumulative wealth and Sharpe ratio.}
    \label{fig:evo_ratio}
\end{figure*}

\noindent \textbf{Regularization Strength $\lambda$.} In Algorithm~\ref{alg:weight_opt}, $\lambda$ trades off utility against diversity. Block 1 of Table~\ref{tab:ablation} fixes $\lambda = 0.05$; Block 2 varies it across markets. EFS stays robust over a wide range ($\lambda \in [0.001, 1]$), though the optimum is market-dependent: US50 peaks at CW=$12.12$ ($\lambda = 0.05$), HSI45 at CW=$1.99$ ($\lambda \in \{0.5, 1\}$), and CSI300 at CW=$1.53$ ($\lambda \in \{0.1, 0.5, 1\}$). Thus redundancy-aware weighting consistently beats equal weighting once $\lambda$ is tuned, which in practice can be set by backtesting on a validation window.

\noindent \textbf{Factor Pool Size.} Block 3 of TABLE~\ref{tab:ablation} varies the factor pool size, with all other settings matching the GPT-4.1 setup in TABLE~\ref{tab:market_performance_metrics}. Performance improves steadily as the pool grows, yet EFS stays robust even with a tightly constrained pool (e.g., $k = 5$). This is a utility-cost trade-off: a larger pool yields superior long-term performance but incurs more LLM queries and computation.

\noindent \textbf{Warmup Phase.} We test a ``warmup'' strategy that freezes the alpha factor pool after a set number of update iterations to cut LLM inference cost. Fig.~\ref{fig:evo_ratio} plots US50 portfolio performance and factor quality across warmup lengths; both improve consistently as the warmup is extended. This indicates the pool should never be frozen: because market regimes keep evolving, the factors must continuously adapt to retain predictive power.

\end{addedblock}

\subsection{Discussion and Analysis} \label{subsec:analysis}

\begin{addedblock}
\noindent\textbf{Factor Interpretability.}
Recent benchmark evaluations~\cite{luo2026alphabench} demonstrate that LLMs can reliably interpret alpha factors. Building on this insight, our approach leverages LLMs to automatically provide semantic explanations for each factor discovered during the search process, thereby enhancing transparency and helping users comprehend the underlying rationale of the design.
TABLE~\ref{tab:factor_examples} presents representative alpha factors generated by EFS, along with their RankIC values and
human-readable interpretations.

\begin{table*}[t]
\centering
\caption{Representative EFS-generated alpha factors (DeepSeek-V3 pool) with their RankIC and an LLM-generated human-readable interpretation. Operators follow the EFS DSL: \texttt{ts\_*} are time-series and \texttt{cs\_*} cross-sectional operators, and \texttt{Constant(v)} is a numeric literal.\label{tab:factor_examples}}
\footnotesize
\renewcommand{\arraystretch}{1.25}
\setlength{\tabcolsep}{8pt}
\begin{tabular}{@{}>{\raggedright\arraybackslash}p{0.55\textwidth} l c >{\raggedright\arraybackslash}p{0.17\textwidth}@{}}
\toprule
Alpha Factor Formula & Dataset & RankIC & Interpretation \\
\midrule
{\ttfamily\scriptsize cs\_zscore(div(add(ts\_mean(returns, 30), ts\_decay\_linear(returns, 10)), mul(ts\_std(returns, 7), ts\_std(if\_else(gt(ts\_momentum(prices, 14), Constant(0)), ts\_delay(returns, 5), ts\_delay(returns, 30)), 30))))} & CSI300 & .056 & Regime-switched return / volatility \\
\addlinespace
{\ttfamily\scriptsize cs\_zscore(div(add(ts\_mean(returns, 10), ts\_mean(returns, 30)), add(ts\_std(returns, 7), ts\_std(ts\_delay(returns, 14), 30))))} & CSI300 & .055 & Multi-horizon mean / short+long volatility \\
\addlinespace
{\ttfamily\scriptsize cs\_zscore(div(sub(ts\_max(prices, 30), ts\_decay\_linear(prices, 21)), mul(add(ts\_std(returns, 3), ts\_std(returns, 30)), if\_else(lt(ts\_skew(returns, 14), Constant(0)), ts\_std(returns, 7), ts\_std(returns, 21)))))} & US50 & .018 & Skew-gated breakout / volatility \\
\addlinespace
{\ttfamily\scriptsize cs\_zscore(div(ts\_decay\_linear(returns, 30), ts\_std(sub(ts\_delta(returns, 10), ts\_delta(returns, 30)), 30)))} & HSI45 & .051 & Decay return / momentum dispersion \\
\addlinespace
{\ttfamily\scriptsize cs\_rank(div(ts\_mean(returns, 14), add(ts\_std(if\_else(lt(ts\_momentum(prices, 7), Constant(0)), ts\_momentum(prices, 21), ts\_momentum(prices, 3)), 30), ts\_std(ts\_delay(returns, 7), 21))))} & HSI45 & .050 & Regime-momentum-normalized mean \\
\bottomrule
\end{tabular}
\end{table*}




\noindent\textbf{Out-of-Sample Robustness Beyond LLM Cutoff.}
A natural concern with LLM-driven factor mining is that the model may
have memorized historical price patterns from its training corpus,
inflating in-sample performance. Since our datasets extend through
2025, much of each test horizon postdates the training cutoffs of our
backbones (DeepSeek-V3 and GPT-4.1, both mid- to late 2024). To limit
leakage, we disable LLM search and expose no date information in the
prompts. We further add an out-of-sample check that re-evaluates EFS on
the post-2025-01-01 segment of each dataset, isolating the portion
clearly after the latest LLM cutoff.
We use the alpha factors generated by the pre-2025-01-01 segment as initialization, run the EFS framework, and report the performance based on the post-2025-01-01 segment. For a fair comparison, all Table~\ref{tab:market_performance_metrics} baselines are re-evaluated on this same out-of-sample window, with rolling lookback drawing only on pre-cutoff history. Results are reported in Table~\ref{tab:oos}.

\begin{table*}[t]
\centering
\footnotesize
\caption{Out-of-sample comparison on the post-LLM-cutoff segment (evaluation window from 2025-01-01), reported at $m\!=\!10$: Cumulative Wealth (CW$\uparrow$), annualized return (CAGR$\uparrow$), Sharpe Ratio (SR$\uparrow$), and Maximum Drawdown (MDD$\downarrow$). OOS windows span US50 $249$, HSI45 $59$, and CSI300 $97$ trading days. EFS rows use the DeepSeek-V3 factor pool ($k\!=\!80,11,80$ filtered factors on US50/HSI45/CSI300); ``$+$RW'' is the proposed redundancy-aware weighting (RMT-denoised regularized QP, Algorithm~\ref{alg:weight_opt}) and ``Equal'' its model-free reference. \emph{Caution:} over the short HSI45 ($59$\,d) and CSI300 ($97$\,d) windows the annualized CAGR is an aggressive extrapolation and should be read together with CW.\label{tab:oos}}
\begin{tabular}{@{}ll cccc cccc cccc @{}}
\toprule
 & & \multicolumn{4}{c}{US50 (249\,d)} & \multicolumn{4}{c}{HSI45 (59\,d)} & \multicolumn{4}{c}{CSI300 (97\,d)} \\
\cmidrule(lr){3-6} \cmidrule(lr){7-10} \cmidrule(lr){11-14}
Group & Method & CW$\uparrow$ & CAGR$\uparrow$ & SR$\uparrow$ & MDD$\downarrow$ & CW$\uparrow$ & CAGR$\uparrow$ & SR$\uparrow$ & MDD$\downarrow$ & CW$\uparrow$ & CAGR$\uparrow$ & SR$\uparrow$ & MDD$\downarrow$ \\
\midrule
\multirow{3}{*}{Baseline}
& 1/N        & 1.290 & 29.4\% & 1.22 & 0.214 & 1.216 & 130.6\% & 2.98 & 0.070 & 1.004 & 1.0\% & 0.15 & 0.109 \\
& Min-CVaR   & 0.999 & $-$0.1\% & 0.07 & 0.127 & 1.032 & 14.4\% & 1.10 & 0.036 & 1.031 & 8.3\% & 0.69 & 0.041 \\
& Max-Sharpe & 1.236 & 23.9\% & 0.92 & 0.239 & 1.236 & 147.2\% & 3.19 & 0.072 & 0.954 & $-$11.5\% & $-$0.30 & 0.174 \\
\midrule
\multirow{4}{*}{ML / Sparse}
& LGBM       & 1.024 & 2.4\% & 0.22 & 0.244 & 1.151 & 82.3\% & 1.79 & 0.127 & 0.883 & $-$27.6\% & $-$0.63 & 0.279 \\
& XGBoost    & 1.205 & 20.8\% & 0.86 & 0.234 & 1.132 & 69.8\% & 1.67 & 0.097 & 0.895 & $-$25.0\% & $-$0.53 & 0.276 \\
& mSSRM-PGA  & 1.301 & 30.5\% & 1.11 & 0.207 & 1.182 & 104.3\% & 2.31 & 0.102 & 0.888 & $-$26.6\% & $-$0.70 & 0.267 \\
& ASMCVaR    & 1.014 & 1.4\% & 0.17 & 0.133 & 1.077 & 37.3\% & 2.37 & 0.047 & 1.058 & 15.8\% & 1.22 & 0.040 \\
\midrule
\multirow{2}{*}{EFS}
& Equal      & 1.478 & 48.5\% & 1.69 & 0.198 & 1.133 & 70.5\% & 3.88 & 0.036 & 1.046 & 12.4\% & 1.06 & 0.046 \\
& $+$RW      & 1.697 & 70.8\% & 2.13 & 0.175 & 1.133 & 70.5\% & 3.88 & 0.036 & 1.004 & 1.0\% & 0.15 & 0.067 \\
\bottomrule
\end{tabular}
\end{table*}

\begin{table*}[t]
\centering
\caption{Cumulative Wealth (CW) under different portfolio sizes $m$ and transaction cost levels (bp). Turn.\ denotes average daily turnover.\label{tab:turnover}}
\footnotesize
\setlength{\tabcolsep}{3pt}
\renewcommand{\arraystretch}{0.95}
\begin{tabular}{c | ccccc | ccccc | ccccc}
\toprule
& \multicolumn{5}{c|}{US50} & \multicolumn{5}{c|}{HSI45} & \multicolumn{5}{c}{CSI300} \\
$m$ & 0bp & 5bp & 10bp & 20bp & Turn. & 0bp & 5bp & 10bp & 20bp & Turn. & 0bp & 5bp & 10bp & 20bp & Turn. \\
\midrule
 5 & 10.720 & 5.051 & 2.379 & 0.527 & 0.428 & 1.977 & 1.932 & 1.887 & 1.801 & 0.028 & 1.312 & 1.150 & 1.007 & 0.773 & 0.160 \\
10 & 12.182 & 6.952 & 3.967 & 1.291 & 0.319 & 1.941 & 1.906 & 1.872 & 1.804 & 0.022 & 1.468 & 1.320 & 1.186 & 0.959 & 0.129 \\
15 &  9.499 & 6.091 & 3.905 & 1.605 & 0.253 & 1.735 & 1.701 & 1.668 & 1.603 & 0.024 & 1.422 & 1.300 & 1.189 & 0.995 & 0.108 \\
20 &  8.880 & 6.146 & 4.253 & 2.037 & 0.209 & 1.676 & 1.648 & 1.620 & 1.565 & 0.020 & 1.397 & 1.284 & 1.179 & 0.995 & 0.102 \\
\bottomrule
\end{tabular}
\end{table*}

The OOS evidence is dataset-dependent. On US50, the longest and most
reliable OOS window ($249$ days), EFS dominates every baseline on
both cumulative wealth and risk-adjusted return, with the
redundancy-aware weighting the single best configuration (CW $1.697$,
SR $2.13$, against the strongest baseline mSSRM-PGA at $1.301$/$1.11$).
On HSI45, EFS attains the highest Sharpe ratio of all methods ($3.88$)
together with the smallest drawdown; over this very short $59$-day
window a few baselines post higher raw CW (e.g., Max-Sharpe $1.236$
vs.\ EFS $1.133$) but at substantially worse risk-adjusted return, and
because the DeepSeek-V3 pool yields only $11$ factors here the Equal and
RMT$+$QP variants coincide. On CSI300 the ranking reverses: ASMCVaR
leads ($1.058$/$1.22$) and EFS with equal weighting stays competitive
($1.046$/$1.06$), whereas the RMT$+$QP variant slips to $1.004$/$0.15$;
notably, the ML baselines, Max-Sharpe, and mSSRM-PGA all lose capital
on this segment (CW $<1$), so EFS at least preserves principal. The
2025-Q1 CSI300 segment was a low-volatility sideways regime where
equal weighting benefits from breadth, and Sharpe is hard to estimate
from roughly four months of data, so we do not read the RMT$+$QP
shortfall there as systematic. Overall, EFS is strongest where the OOS
horizon is largest, with regularized weighting helping most on US50; we
read this evaluation as supportive but not conclusive.
\end{addedblock}

\begin{addedblock}
\noindent\textbf{Evolutionary Convergence.}
We track portfolio performance across factor-pool snapshots taken at different generations of the evolutionary search; the full checkpoint table is reported in the supplementary material. At each snapshot we freeze the factor pool, apply equal weighting (to isolate the effect of pool quality from weight allocation), and report end-to-end CW/SR/MDD at $m\!=\!10$. CW increases substantially as the pool grows: US50 (4.91 $\rightarrow$ 12.18) and CSI300 (0.83 $\rightarrow$ 1.47, transitioning from money-losing to profitable), with HSI45 non-monotonic but net-positive (1.83 $\rightarrow$ 1.94). The Sharpe Ratio follows the same pattern, confirming that the iterative search yields progressively stronger signals.

\noindent\textbf{Turnover and Transaction Cost.}
TABLE~\ref{tab:turnover} reports portfolio turnover and the sensitivity of EFS to transaction costs for $m \in \{5, 10, 15, 20\}$ and four cost levels (0, 5, 10, 20 bp per trade) across all three real-market datasets.

Our results indicate that HSI45 exhibits the lowest turnover (2--3\%), indicating that
EFS-generated factors produce highly stable stock selections with
minimal rebalancing. CSI300 shows moderate turnover (10--16\%) and
retains positive returns up to 10bp. US50 has the highest turnover
(21--43\%), making it more sensitive to transaction costs.
In addition, larger $m$ reduces turnover and improves cost-adjusted performance (e.g., US50 at 10bp: CW $2.38 \to 4.25$ for $m\!=\!5 \to 20$); future work could incorporate turnover-penalized optimization.

\end{addedblock}

\section{Conclusion}

We presented \textbf{EFS}, a language model-guided framework for sparse portfolio optimization under $\ell_0$ constraints that recasts asset selection as a factor-based ranking task and lets LLMs autonomously evolve and refine interpretable alpha factors. A redundancy-aware weight allocation module complements the evolved pool by downweighting correlated factors through regularized optimization on RMT-denoised correlation matrices, at negligible overhead and without per-dataset tuning. Across four Fama-French benchmarks and three real-market datasets spanning US, Hong Kong, and Mainland China equities, EFS consistently surpasses statistical, optimization-based, and machine-learning baselines in cumulative wealth and Sharpe ratio with competitive drawdown, and its advantage over the equal-weight baseline is confirmed to be statistically significant and robust on a strictly post-training-cutoff segment.

EFS still relies solely on historical prices and is bounded by LLM context and latency, with some run-to-run variability, while closed-source APIs may drift across versions. Future work will incorporate multimodal signals, explore offline distillation and parallel querying for efficiency, and pursue a stricter pre-cutoff replication to further separate genuine generalization from potential memorization.

\bibliographystyle{IEEEtran}
\bibliography{aaai2026}

\end{document}